\documentclass[twocolumn,showpacs,preprintnumbers,amsmath,amssymb]{revtex4}
\usepackage{graphicx}
\usepackage{dcolumn}
\usepackage{bm}

\begin{document}

\title{Characterizing micro-macro transitions with an atomic-vapor-based linear optical amplifier}

\author{Zhifan Zhou$^1$}
\email{zfzhoucom@gmail.com}
\altaffiliation[Present address: ]{Department of Physics, Ben-Gurion University of the Negev, Beersheba 84105, Israel. }
\author{Ulrich Vogl$^2$}
\author{Ryan T. Glasser$^3$}
\author{Zhongzhong Qin$^1$}
\author{Yami Fang$^1$}
\author{Jietai Jing$^1$}
\email{jtjing@phy.ecnu.edu.cn}
\author{Weiping Zhang$^1$}
\email{wpzhang@phy.ecnu.edu.cn}

\affiliation{$^1$ Quantum Institute for Light and Atoms, State Key Laboratory of Precision Spectroscopy, Department of Physics, East China Normal University, Shanghai 200062, P. R. China \\$^2$ Max Planck Institute for the Science of Light, G$\ddot{u}$nther-Scharowsky-Strasse 1, Bau 24, 91058 Erlangen, Germany \\ $^3$ Department of Physics and Engineering Physics, Tulane University,
6400 Freret Street, New Orleans, Louisiana 70118, USA}

\begin{abstract}
Fundamentally, the dynamics of micro-macro transitions is instrumental to understanding the process of quantum-to-classical transitions; technologically, it can also facilitate the detection of the microscopic signals in quantum experiments via convenient detectors. Here, we demonstrate a scheme to characterize micro-macro transitions based on a four-wave mixing linear optical amplification process in a hot rubidium vapor. The linear optical amplifier provides a large optical gain of $10^7$ for injected single-photon-level pulses, enabling photon-number-resolving detection by average via non-single-photon counting detectors with a large dynamic range. The scheme exhibits strong dispersion which is sensitive to the input's change at the single-photon level, resulting in the group-velocity delay time scaling with $1/\sqrt{N}$, where $N$ is the average input photon number. The output probe and conjugate modes have different coefficients of this $1/\sqrt{N}$ scaling, indicating the coefficient can serve as an efficient parameter to characterize the specified micro-macro transitions. The demonstrated results are generally applicable for quantum detection and optical signal processing in light-atom interfaces. Furthermore, the present system is suitable for the study of relevant time-resolved dynamics of the quantum-to-classical transitions. 
\end{abstract}


\maketitle

\begin{figure}[tbhp]
\centering
\includegraphics[width=11.5cm,height=9.5cm]{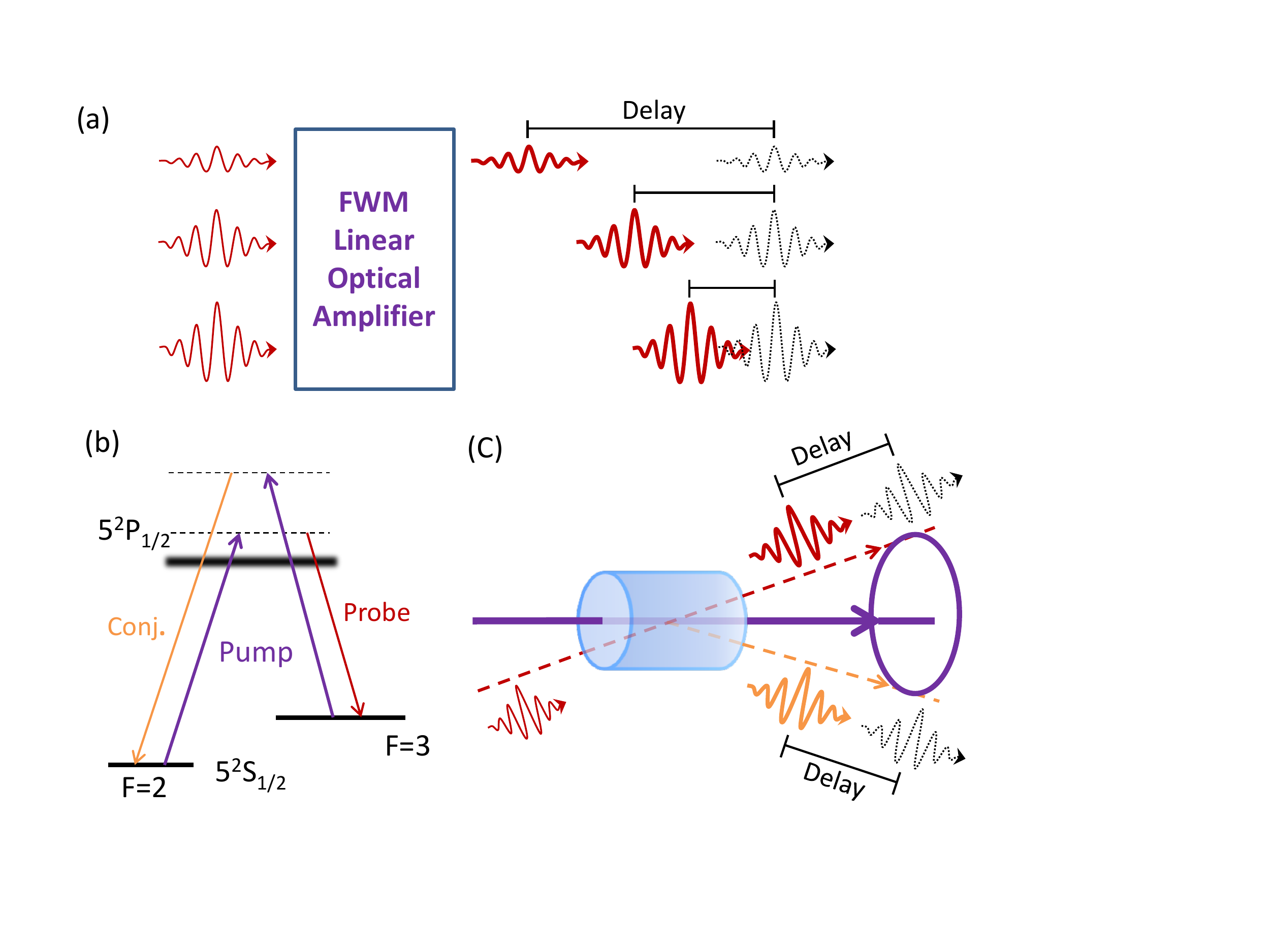}
\caption{Experimental scheme. (a), The linear optical amplifier amplifies the microscopic-level input into the macroscopic regime and the transition time, which is indicated by the group-velocity delay time, is observed to be differentiated with the increased input at the microscopic level. (b), Atomic energy levels of the D1 transition of Rb$^{85}$ involved in the four-wave mixing\,(FWM) process based on the double-lambda configuration. FWM light fields can be coupled through the atomic coherence between the two ground states, which reduces the resonant losses and increases the relevant nonlinearity. Combined with the large detuning relative to the Doppler-broadened spectrum in a hot Rb$^{85}$ vapor, resonant absorption and spontaneous emission are suppressed. With a strong, non-depleted pump light beam, the system behaves like a linear optical amplifier for the input photons. Furthermore, the quantum interference via the coupling between the light fields and the atomic coherence results in a large group-velocity delay. (c), The geometry of the experiment. The amplifier provides a large optical gain of $10^7$ to amplify the input single-photon-level signals into the macroscopic level, and to generate a conjugate beam on the opposite side. }
\label{}
\end{figure}

Optical amplification has played an important role in optical signal processing and optical communication\,\cite{Loock, Girvin}. The scheme of amplifying the single-photon qubit via stimulated emission is also proposed to facilitate the detection in quantum experiments with convenient detectors\,\cite{Gisin}.  On the other hand, the dynamics of optical amplification is instrumental to understanding the process of quantum-to-classical transitions\,\cite{Stim, Howell2002} as the fundamental quantum non-cloning law imposes the impossibility of perfectly copying an unknown arbitrary quantum state\,\cite{Caves}.  The connection between micro-to-macro transitions and quantum-to-classical transitions has been investigated with optical parametric amplifiers in crystal-systems\,\cite{Crystal2008, MartiniRMP} and with four-wave mixing in fiber-systems\,\cite{Peer2015}. However, due to the large inhomogeneous broadening in the noted condensed-matter materials, the dispersion property and the micro-to-macro transition-time are not well-resolved\,\cite{Crystal2008, MartiniRMP, Peer2015}. Note there is another aspect of micro-macro state is prepared by mixing the micro and macro state via beam splitters\,\cite{Lvovsky2, Gisin2}.  

Alkali-metal atomic vapor, with well-defined energy level structures, high feasible optical depths and much smaller inhomogeneous broadening, allows for quantum interference between different channels, e.g. electromagnetically induced transparency\,(EIT)\,\cite{Harris1989, Harris1991, Harris1995, Hau, LukinReview, EisamanNature, EIT, CesiumSlow}, and enhanced resonant light-atom interactions due to the radiative interference and the induced atomic coherence\,\cite{LukinReview, EIT, Lukin1998, Lukin2000, PolzikReview}. Theoretical research has pointed out the efficient parametric amplification for the infinitesimally small initial values is possible in a resonantly driven multilevel atomic system\,\cite{Lukin1998}. EIT based on quantum interference can render an otherwise opaque medium transparent over a narrow spectral range. The resulting steep dispersion and the group-velocity delay are used to buffer optical information and interrogate the dynamics of the light-matter interactions\,\cite{Hau, LukinReview, EisamanNature, EIT, CesiumSlow, KashSlow, FWMSlow, HowellFWM, AlbertoNature, HybridSlow, Cavity, VIT}. Up to now, the demonstrated slow-light phenomena are tunable by the macroscopic means such as by changing the strong pump power or by changing the medium temperature\,\cite{Hau, LukinReview, EisamanNature, EIT, CesiumSlow,KashSlow,FWMSlow,HowellFWM,AlbertoNature,HybridSlow}, but are not sensitive to the microscopic input's change at the single-photon level\,(which is theoretically proposed in \cite{Cavity} and experimentally explored towards realization in \cite{VIT}, both with cavity systems).  It is then complementary to have the access to the information from the microscopic view.

In this work, to characterize the micro-macro transitions with an atomic-vapor-based linear optical amplifier\,(LOA), the four-wave mixing\,(FWM) process based on the double-lambda configuration\,(Fig. 1(b)) in a hot rubidium vapor\,\cite{FWMPRA2008, BoyerScience} is used. In this process, two pump photons are converted into one probe photon and one conjugate photon in each cycle such that the atomic states are preserved after each cycle. The light fields can be coupled via the atomic coherence between the two hyperfine ground states, which allows for reducing the resonant losses and increasing the relevant nonlinearities \cite{LukinReview, EIT, Lukin1998, Lukin2000, PolzikReview}.  Combined with the large detuning relative to the Doppler-broadened spectrum in a hot Rb$^{85}$ vapor, resonant absorption and spontaneous emission are suppressed. With a strong pump light beam, which ensures the condition of low pump depletion, the system behaves like a linear optical amplifier for the probe and the conjugate. Furthermore, the quantum interference via the coupling between the light fields and the atomic coherence results in a large group-velocity delay\, \cite{EIT, FWMSlow, AlbertoNature}. Here, we make use of the FWM-based LOA to probe the micro-macro transition's dynamics, involving a large dynamic range of $~70$\,dB in photon flux.  

\begin{figure*}
\begin{center}
\includegraphics[width=14.5cm,height=12cm]{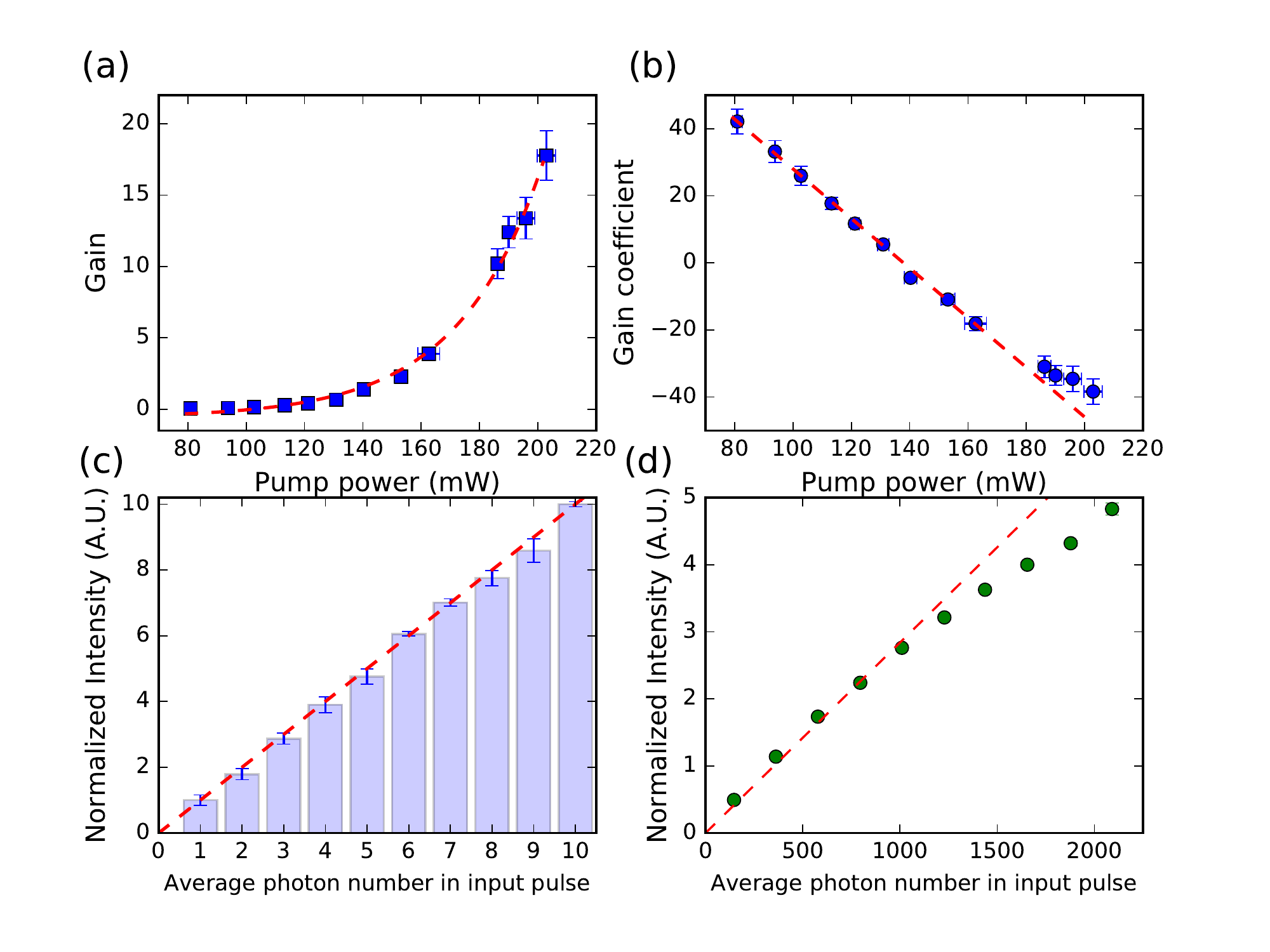}
\end{center}
\caption{Gain properties of the linear optical amplifier. (a), Gain versus pump power. The gain is exponentially\,(guide to eye, the red dashed curve) increasing with the pump power. (b), The gain coefficient versus pump power. It can be seen that the gain coefficient is linearly\,(guide to eye, the red dashed curve) tunable in a large range depending on the pump power. (c), The output amplitude of the conjugate mode versus the average photon number in the input pulse. The red line is linear with a slope of one, and is used as a guide for the eye. To get the significant signal to noise ratios, the plots displayed have been averaged by $10^5$ times for the lowest plot and $10^4$ times for the highest plot. (d), The large scale of dynamic range of conjugate versus average photon number in input pulse. The linearity\,(guide to eye, the red dashed curve) is preserved up to $\sim$1000 average photon number. Error bars represent one standard deviation, combined statistical and systematic uncertainties. Error bars in (d) are smaller than the symbols. }
\label{}
\end{figure*}
A high-gain FWM-based LOA requires a high effective optical depth, such that a 75-mm-long rubidium cell\,(Fig. 1(c)) with a high atomic density\, ($2\times10^{13}$ atoms/cm$^3$) at a temperature of 130$^{\circ}$C is used. We begin by characterizing the FWM-based amplification process by seeding a bright coherent beam\,(27\,$\mu$W) with a diameter of 1\,mm at the near-resonance probe frequency and measuring the output's dependence on the power of the pump beam\,(with a diameter of 2\,mm). The gain is measured as the ratio between the output conjugate power and the input seed. The gain of FWM, $G=e^{-g\cdot L}$, where {\it g} is the gain coefficient and {\it L} is the length of the rubidium cell, is found to exponentially increase with the pump power\,(Fig. 2(a)). Correspondingly, the gain coefficient is then found to be linearly decreasing with pump power\,(Fig. 2(b)). At a pump power of 130\,mW, the gain coefficient is 0, which indicates that the gain is unity. The gain coefficient keeps decreasing linearly until saturation around 180\,mW. It is worth noting that the gain for a low-light-level seed may be much higher than a high-light-level seed's gain\,(for example, 27\,$\mu$W here) in the amplification process\,\cite{GauthierPRA}, due to the limited pump power and the saturation of the gain medium.

We operate here in the high-gain regime to demonstrate the LOA for the single-photon-level input with a pump power of 300\,mW, as the gain-saturated pump power is higher for the low-light-level input\,(compared with the previous 180\,mW for a bright seed of 27\,$\mu$W). When the input probe and conjugate modes are vacua, the strong coupling between the pump field and the atoms in FWM can provide large gain for the spontaneously emitted probe and conjugate photons in a single-pass configuration. Conical emission is then generated with an output pattern of a ring with the photons at the probe and conjugate frequencies\,(see the Supplementary Information). The azimuthal angle between the pump beam and the ring is 8\,mrad, indicating the optimal orientation to fulfill the phase-matching condition in this FWM scheme\,\cite{PhaseMatch}. As the gain is high, a single-photon-level injected probe field along the azimuthal angle can trigger the stimulated FWM process, amplifying the input probe and generating the conjugate\,\cite{Lukin1998, GauthierScience}.

According to Fig.\,1(b), the probe photons are more near-resonant than the conjugate photons, and are used to seed the process. By injecting a weak probe beam, stimulated FWM occurs in which the probe is amplified by a factor of 10$^7$ and a conjugate is generated on the opposite side of the pump\,(Fig.\,1(c)). We investigate the output power of the conjugate, which suffers less from the Doppler-broadened absorption, as a function of average input probe photon number from 1 to 10 photons\,(Fig.\,2(c)) and for a large scale\,(Fig.\,2(d)), showing the linear amplifications for a large dynamic range of low-light-level inputs. As the outputs are at the macroscopic level, the output signals are detected by off-the-shelf linear detectors\,(non-single-photon counting detectors, e.g. Thorlabs PDB450A). The results demonstrate that the FWM-based LOA is capable of operating at the single-photon level and the linearity is preserved well up to $\sim$1000 average input photon number. The signals observed on the oscilloscope are averaged for $10^5$ times for the lowest plot in Fig.\,2(c) and $10^4$ times for the highest plot in Fig. 2(c) in order to achieve high signal-to-noise ratios\,(SNRs). 

In our experiment, the large gain resulted from high effective optical depth leads to the generation of the conical emission and the partial re-absorption for the probe and pump beams in the Doppler-broadened wings of rubidium vapor\,(see the Supplementary Information), resulting in significant extra noise, which limits the SNR of our system. It is anticipated that working in a regime of lower optical depth to decrease the extra noise, and cascading the system to compensate for the decreased gain, will largely enhance the overall SNR\,\cite{Cascaded}. In the other hand, quantum mechanics predicts that a certain level of quantum noise must be added in any optical amplifier\,\cite{Caves}. When the FWM-based LOA’s gain is tuned into an appropriate level to reveal the intrinsic quantum correlation between the probe and the conjugate in this FWM scheme\,\cite{FWMPRA2008, BoyerScience}, the cascaded FWM system would allow for cancellation of internal quantum noise of an optical amplifier through quantum destructive interference\,\cite{Ou}. 

\begin{figure}
\begin{center}
\includegraphics[width=9.5cm,height=9cm]{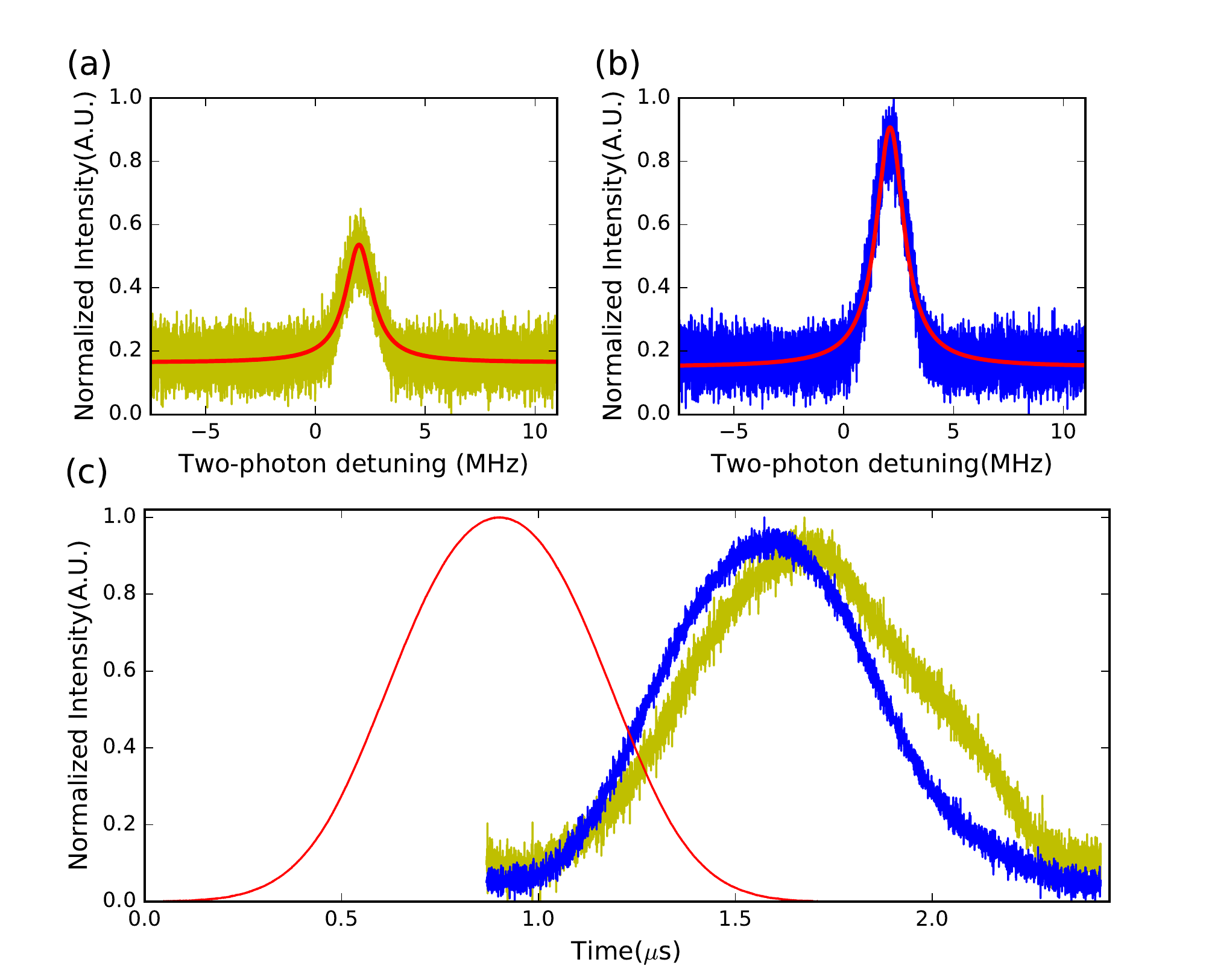}
\end{center}
\caption{Slow light of the output modes. (a), and (b), The gain profiles of the probe and the conjugate versus the two-photon detuning when seeded with a 1\,pW continuous-wave probe beam. Each profile is fitted by a Lorentzian function, indicating the much narrower bandwidths around 1.5\,MHz than the natural bandwidth of Rb$^{85}$, 5.75\,MHz. The output probe beam is weaker than the conjugate due to the Doppler-broadened absorption in its vicinity, which also leads to more distorted output probe pulses. (c), The delays of the output probe and conjugate pulses at two different frequencies when injected with probe pulse containing 0.7\,photons on average. The delay times for the probe and conjugate pulses are 672ns\,(yellow curve) and 592ns\,(blue curve), with the fractional delays of 1.14 and 1.01, respectively. The plots are averaged for $10^4$ times. The red curve is the reference pulse taken with propagation at the vacuum speed of light. All the plots are superposed on a constant background from the unseeded conical emission and normalized to facilitate the comparison of the input and output pulse-shapes.}
\label{}
\end{figure}
In order to characterize the dispersion properties of the FWM process, the probe frequency is scanned across the two-photon resonance\,(with the pump shown in Fig.\,1(b)), revealing two nearly Lorentzian gain profiles at the probe and conjugate frequencies with full width half maximum\,(FWHM) linewidths of 1.48\,MHz and 1.53\,MHz, respectively, as shown in Fig.\,3(a) and Fig.\,3(b). Both of the two gain profiles are much narrower than the natural linewidth of Rb$^{85}$, 5.75\,MHz, due to the quantum interference based on the atomic coherence in the double-lambda configuration\,\cite{FWMSlow,HowellFWM,LukinScience2003}. The amplitude of the output probe mode is smaller than that of the conjugate mode due to the additional Doppler-broadened absorption at the probe frequency\,(Supplementary Information). 

The steep dispersions resulted from the narrow gain profiles give rise to the slow group velocities for the output probe and conjugate signals\,\cite{EIT, Lukin2000, Gorshkov,BoydGauthier}, when tuned to the gain-line center. In Fig.\,3(c), the generated group-velocity delays are displayed with the injected probe pulse containing 0.7\,photons on average. The reference pulse is obtained by measuring the injected strong probe pulses without attenuation when the pump beam is blocked and the probe frequency is tuned far away from the atomic resonances. The group-velocity delays of the probe and conjugate pulses are 672\,ns and 592\,ns, with the fractional delays of 1.14 and 1.01, respectively\,(Fig. 3(c)).  One noted feature is that the conjugate pulse is faster than the probe pulse, which is a well-understood property of the dynamics of FWM process based on the double-lambda configuration\,\cite{FWMSlow,HowellFWM,LukinScience2003}.

\begin{figure*}
\begin{center}
\includegraphics[width=13cm, height=16cm]{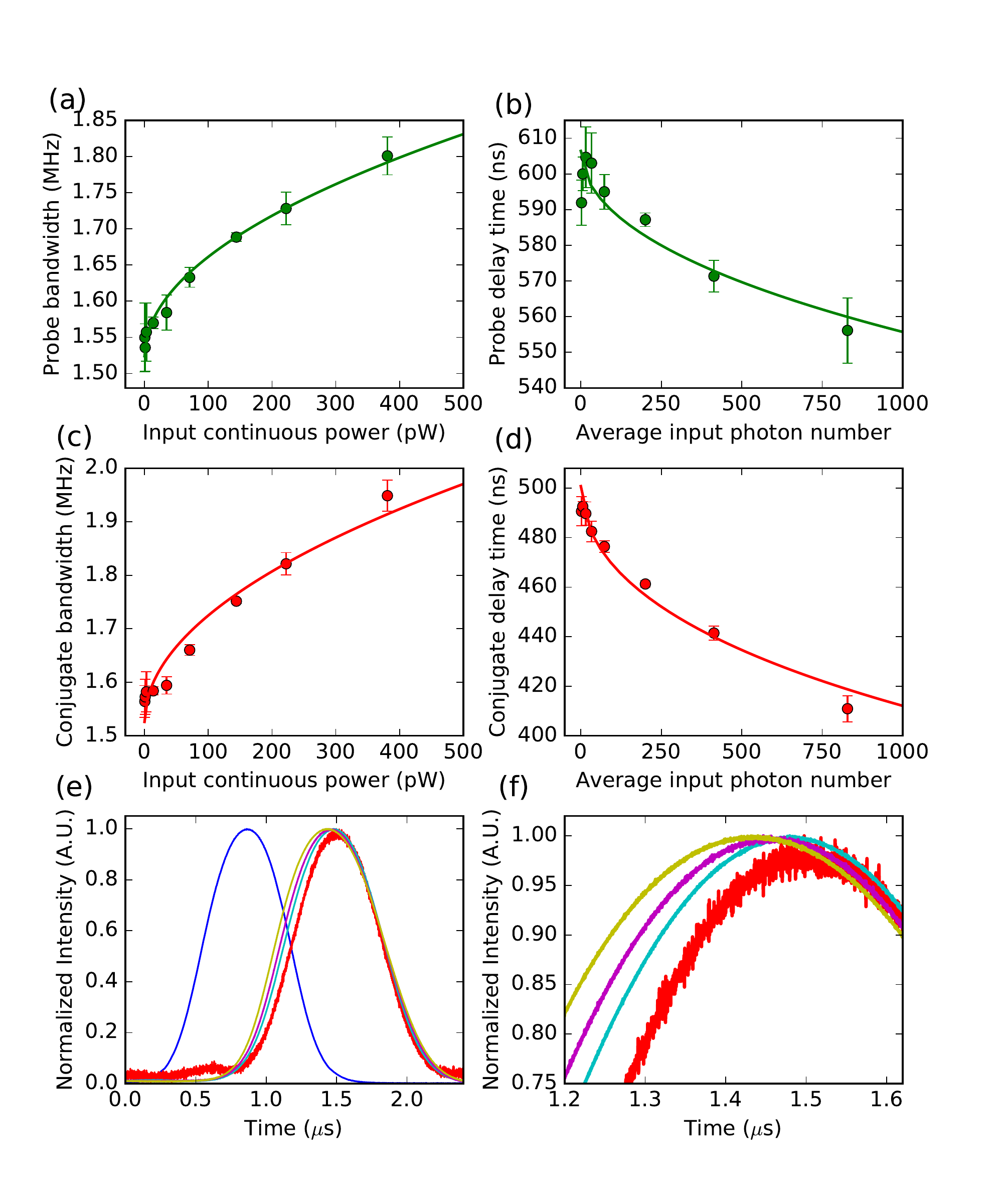}
\end{center}
\caption{ Dispersion-delay time matching. (a), and (c), The bandwidths of the probe and conjugate gain profiles versus the input continuous power. (b), and (d), The delay of the conjugate pulse versus the average input photon number. The fits from Eq.(1) are essentially of the forms of $s\cdot \sqrt{x}+z$ for (a) and (c), and $1/(s\cdot \sqrt{x}+z)$ for (b) and (d), in which $s$ characterizes the dispersion's sensitivity on the input power and $z$ is the off-set determined by the system. For conjugate, $s_c$ and $s_d$ are 1.988 $\pm$0.39$ \times$$10^5 GHz/\sqrt{pW}$ and 2.154 $\pm$0.33$ \times$$10^5 GHz/\sqrt{pW}$. For probe, $s_a$ and $s_b$ are 1.373 $\pm$0.144$ \times$$10^5 GHz/\sqrt{pW}$ and 0.729 $\pm$0.292$\times$$10^5 GHz/\sqrt{pW}$. The conjugate has stronger sensitivity and better bandwidth-delay time matching due to the less absorption. (e-f), Typical delayed pulses and highlighted peak parts in the conjugate mode for different average input photon numbers of 413\,(yellow), 200\,(purple), 73\,(green), 2\,(red), and reference pulse\,(blue). The pulses shown in Fig.\,4(e-f) have the background subtracted and are normalized to assist in the visualization.} 
\label{}
\end{figure*}
The strong coupling at the low-light-level injection allows for investigating the dependence of the dispersion and the group-velocity delay on the average input photon numbers, which is shown in Fig.\,4. Due to the strong coupling, the stronger input continuous power will make the bandwidths of the gain profiles broader\,(Fig.4(a) and Fig.4(c)). To describe this effect, we utilize the Raman gain model based on atomic coherence\,\cite{Lukin2000, Gorshkov}: 
  \begin{equation}
 \Gamma_{gain}=\frac{{\eta}\cdot{\Omega_{pump}}\cdot{\Omega_{probe}}}{\Delta_{Raman}},
\label{eq:initial}
\end{equation}                       
involving the gain profile's bandwidth, $ \Gamma_{gain}$,  the Rabi frequencies of pump and probe, $\Omega_{pump}$ and $\Omega_{probe}$, and the coupling strength, $\eta$. The broadening of the gain profile will lead to the less steep dispersion and result in a shorter group-velocity delay of the pulses, which is characterized as $\tau_d=1/\Gamma_{gain}$\,\cite{Lukin2000}. 

Here we are interested in characterizing the relationship between the dispersion and the resulting delay time in the linear amplification regime indicated in Fig.\,2(d). The results of the gain bandwidth when the input continuous probe power increases from 0.5\,pW to 400\,pW are shown in Fig.\,4(a) and Fig.\,4(c) for the probe and the conjugate, respectively. With the pulse lengths of 587\,ns, it corresponds to the range for the input average photon number from $\sim$1 to $\sim$800 in Fig.\,4(b) and Fig.\,4(d), well within the linear amplification regime\,(Fig.\,2(d)). The resulting change in the bandwidth and the delay time agree quantitatively with Eq.\,(1) over the large dynamic range. The typical delayed pulses in the conjugate mode for different average input photon numbers are shown in Fig.\,4(e). While increasing the input power causes other nonlinearities\,\cite{BoydGauthier, BoydBook} which are responsible for the slight pulse distortions, the pulse peaks\,(Fig.\,4(f)) are depending on the average input photon numbers as predicted by Eq.\,(1). The pulses shown in Fig.\,4(e-f) have the background subtracted and are normalized to assist in the visualization. 

When the group-velocity delay time is tuned by the strong pump beam or the medium temperature\,\cite{Hau, LukinReview, EisamanNature, EIT, CesiumSlow,KashSlow,FWMSlow,HowellFWM,AlbertoNature,HybridSlow}, it is possible to have the same absolute delay time with different pump powers of medium temperatures by the temperature/pump power balancing. It is then difficult to use one absolute delay time for one specific input state to fully characterize all the aspects of the micro-macro transitions.  It will be complementary to have the access to the collective  delay-time behavior of different microscopic inputs, i.e., the relevant coefficient for the delay-time scaling versus the microscopic inputs may be different in different scenarios. In the current work, the output probe and conjugate modes have different coefficients of this $1/\sqrt{N}$ scaling\,(the probe is with 0.729 $\pm$0.292$\times$$10^5 GHz/\sqrt{pW}$, and the conjugate is with 2.154 $\pm$0.33$ \times$$10^5 GHz/\sqrt{pW}$) due to the different absorptions, indicating the coefficient can serve as an efficient parameter to characterize the specified micro-macro transitions.

In conclusion, we have demonstrated a high-gain linear optical amplifier converting the microscopic-level inputs into the macroscopic level based on the four-wave mixing process in a hot rubidium vapor. The gain for different inputs at the single-photon level is identical in the linear amplification regime, enabling photon-number-resolving detection by average via non-single-photon counting detectors with a large dynamic range. The temporal dynamics for the micro-macro transitions is resolved in this highly dispersive medium, showing the group-velocity delay time scaling with $1/\sqrt{N}$, where $N$ is the average input photon number. The experimental results have shown good dispersion-delay time matching in a large dynamic range. 

Our demonstration is generally applicable for quantum detection of photon-number information in atom-based quantum metrology and quantum information\,\cite{PhotonCounting, Migdall}, i.e. it can be used to determine the mean photon number inside the optical SU(1,1) interferometer\,\cite{Florian, PumpUp} to evaluate the ultimate obtainable sensitivity as it is done in the atomic counterpart\,\cite{Markus}. The next intriguing task is to measure the noise figure\,(NF) with different inputs at the microscopic level, which is the SNR of the amplified signal divided by the input signal's SNR. The relation between the different micro-to-macro transition times in the quantized photon-number regime and the quantum noise property will enable new prospects for the study of quantum-to-classical transition\,\cite{Ionization, Bloch}.  It is also anticipated to inject the micro-to-macro transitions with one half of the quantum correlated photon pairs\,\cite{Lvovsky} to interrogate the micro-macro entanglement in the highly dispersive medium. 

This work is supported by National Key Research Program of China under Grant No. 2016YFA0302000 and Grant No. 2011CB921604, National Basic Research Program of China (2016YFA0302103), Natural Science Foundation of China under Grants Nos. 11474095, 11274118, 11234003, 11129402,  91436211, 11374104 and 10974057, the Fundamental Research Funds for the Central Universities,  Program of Introducing Talents of Discipline to Universities (B12024), the SRFDP (20130076110011), the Program for Professor of Special Appointment 
astern Scholar) at Shanghai Institutions of Higher Learning, the Program for New Century Excellent Talents in University (NCET-10-0383), the Shu Guang project supported by Shanghai Municipal Education Commission and Shanghai Education Development Foundation (11SG26), the Shanghai Pujiang Program under Grant No. 09PJ1404400, the Scientific Research Foundation of the Returned Overseas Chinese Scholars, State Education Ministry, and Program of State Key Laboratory of Advanced Optical Communication Systems and Networks (2016GZKF0JT003).


\pagebreak
\newpage

{\bf Supplementary Information}

\renewcommand{\theequation}{S\arabic{equation}}
\setcounter{figure}{0}    
\renewcommand{\thefigure}{S{\arabic{figure}}}

\section{Experimental methods}

The laser beams used in the experiment are derived from a Ti:Sapphire laser\,(Matisse, Spectra-Physics) blue-detuned roughly 1.25\,GHz from the 5S$_{1/2}$\,(F=2) to 5P$_{1/2}$ transition in $^{85}$Rb. The pump is linearly polarized and is spatially filtered using a single-mode optical fiber. The injected probe is derived from one part of the pump and down-shifted by approximately 3.036\,GHz by double-passing an acoustic-optical modulator. Afterwards the probe is spatially filtered using another single-mode optical fiber. The pump and probe beams are perpendicularly polarized and combined with a Glan-Taylor polarizer which has an extinction ratio of 10$^{5}$:1. A naturally abundant rubidium vapor cell with a length of 7.5\,cm is heated to 130$^{\circ}$C as the working temperature, which corresponds to a $^{85}$Rb atomic number density of $2\times10^{13}$  atoms/cm$^3$. The probe pulses are highly attenuated down to the single-photon level with neutral density filters before being sent into the rubidium cell. The average photon number per pulse is determined by the measured peak power in the range of pW and a pulse width of 587 ns. The output pulses are measured by off-the-shelf photodetectors from Thorlabs\,(PDB450A). To observe the signals with high vertical resolution, we used a 12-bit analog-to-digital converter oscilloscope\,(Teledyne Lecroy HD 4096). To observe the signals with high horizontal resolution, we used a high bandwidth\,(2.5GHz) oscilloscope\,(Tektronix 7254C). Error bars shown in the paper represent one standard deviation, combined statistical and systematic uncertainties.
\begin{figure}
\centering
\includegraphics[width=14.5cm,height=11.5cm]{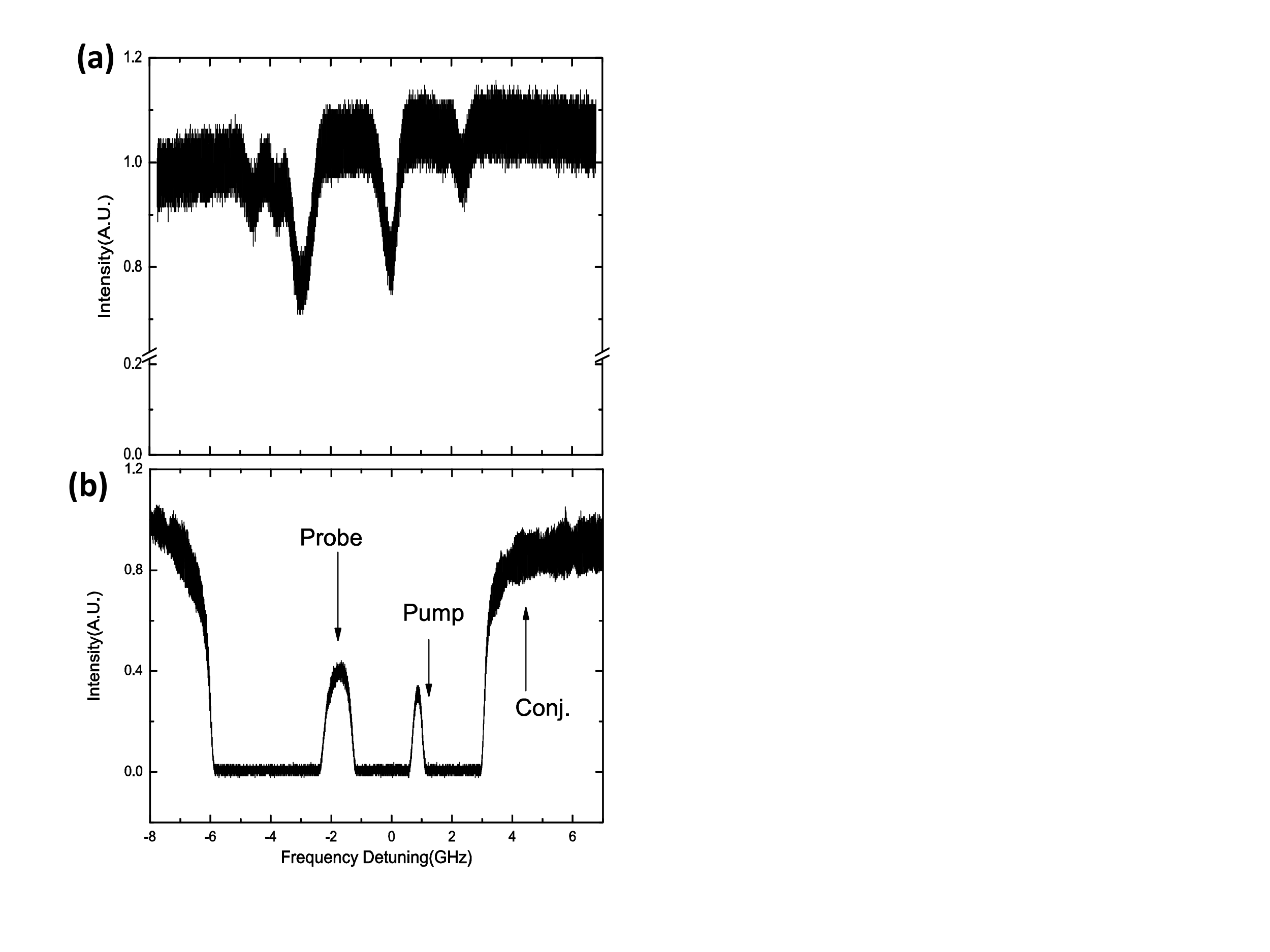}
\caption{ Doppler-broadened spectra of the rubidium cell at the room temperature {\bf (a),} and at the working temperature of 130$^{\circ}$C  {\bf (b).} The frequencies of probe, pump and conjugate are indicated by the arrows.}
\label{}
\end{figure}

\begin{figure*}[tbph]
\centering
\includegraphics[width=10.5cm,height=12.5cm]{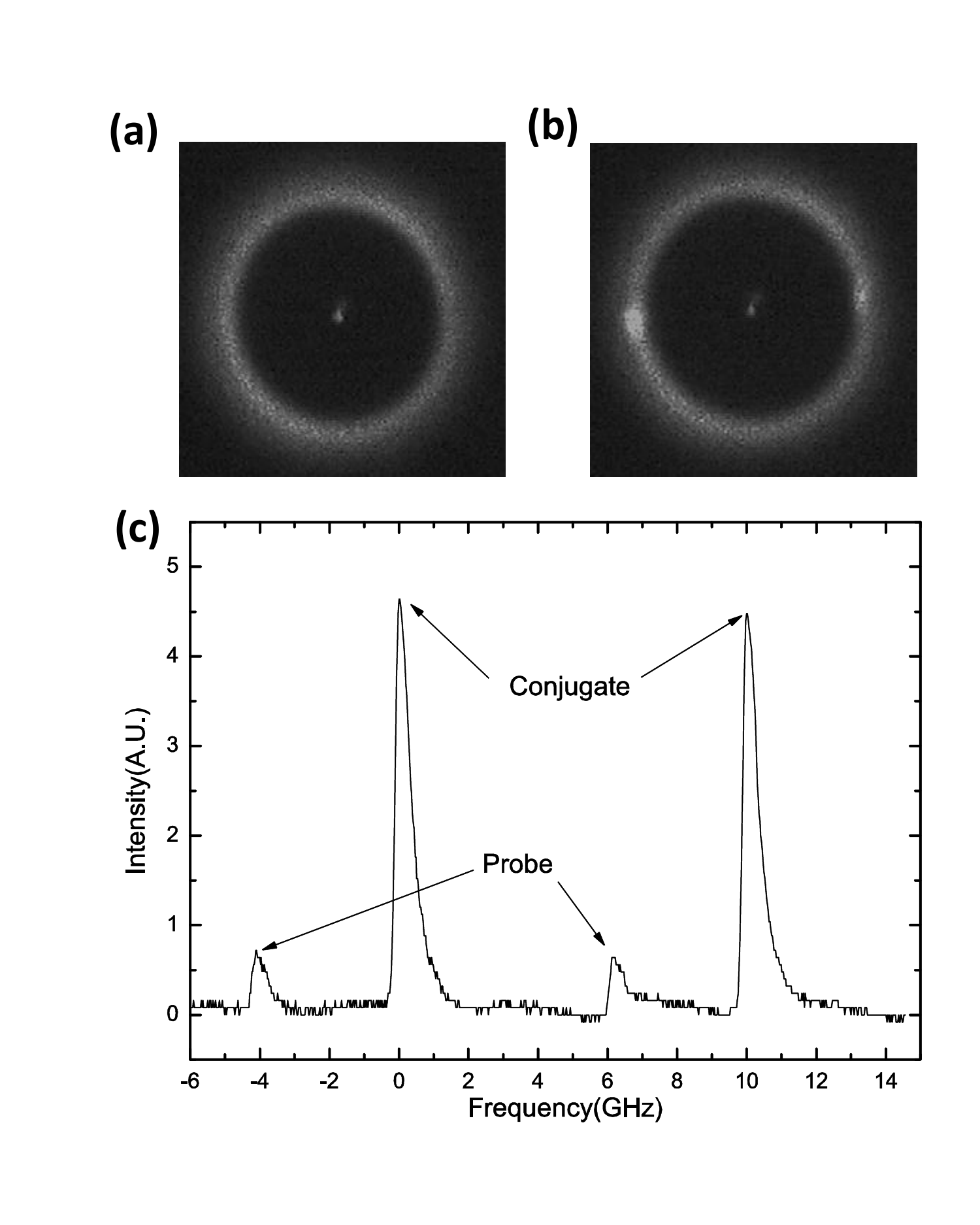}
\caption{ FWM output pattern observed in the far field for the case without input {\bf (a),} and with 1pW continuous input {\bf (b)}. The bright spots are the amplified probe beam\,(right spot) and the generated conjugate beam\,(left spot), respectively. {\bf (c),} The spectrum of conical emission photons is measured by a Fabry-Perot cavity with a free spectral range of 10\,GHz. The displayed figure contains the result of two scanning cycles.  }
%
\includegraphics[width=9.2cm,height=8.5cm]{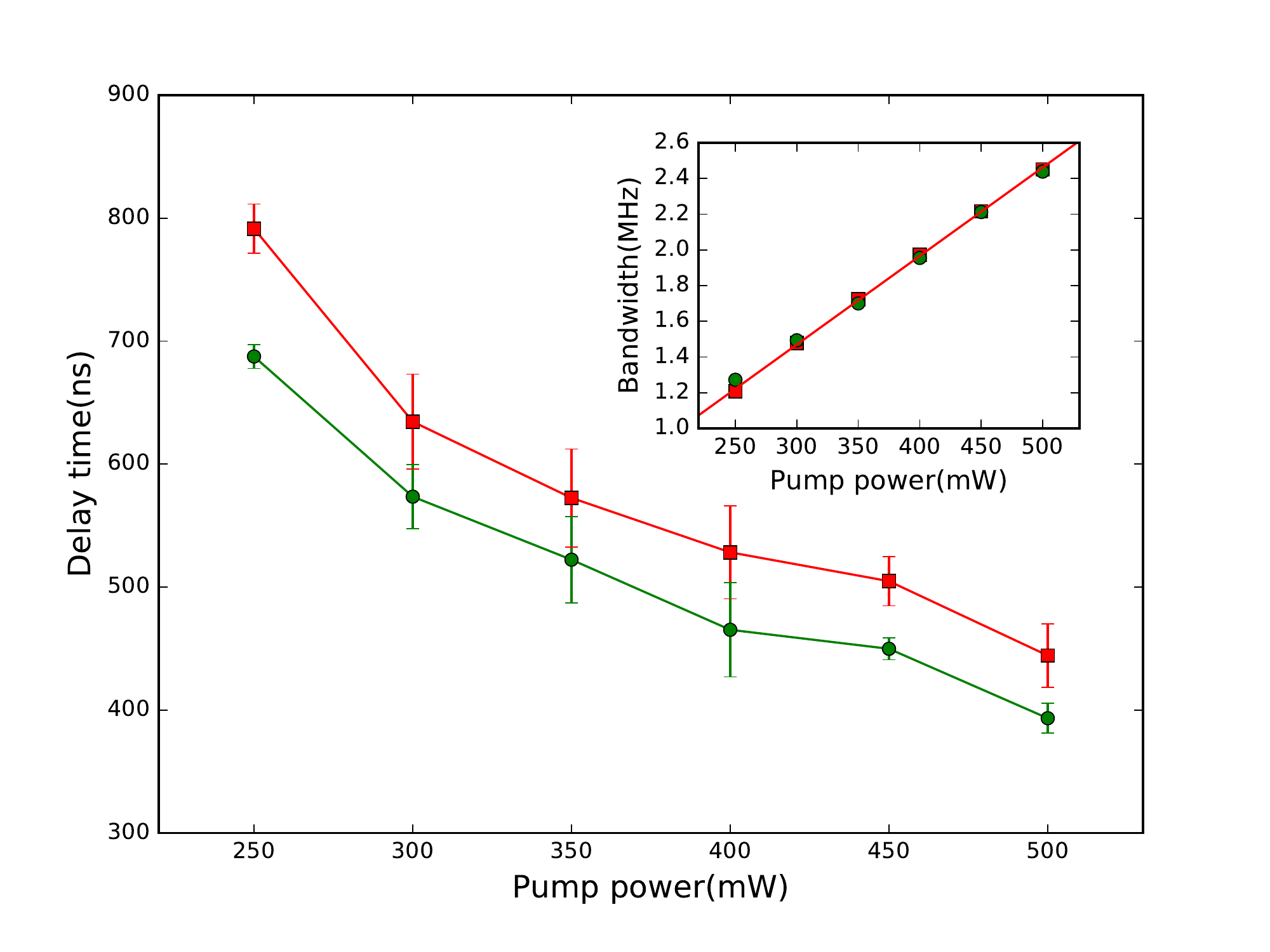}
\caption{The delays of the output probe pulses\,(red squares) and the output conjugate pulses (green circles) versus pump power are shown. The upper right inset shows the gain profile's bandwidth of the probe\,(red squares) and conjugate\,(green circles) versus pump power. Due to the power-broadening effect, the bandwidth is proportional to the pump power, and thus shows good agreement with a linear fit. This bandwidth broadening makes the corresponding dispersion less steep and leads to the less group-velocity delay. Error bars represent one standard deviation, combined statistical and systematic uncertainties. The error bars in the inset figure are smaller than the symbols' size. }
\label{}
\end{figure*}

\section{Doppler-broadened absorption, conical emission and spectra}

The Doppler-broadened spectra of the rubidium cell are shown in Fig.\,S1. The origin corresponds to the 5S$_{1/2}$ (F=2) to 5P$_{1/2}$ transition of $^{85}$Rb D1 line.  We set the pump frequency around 1.25\,GHz blue-detuned from the 5S$_{1/2}$ (F=2) to 5P$_{1/2}$ transition to avoid the Doppler-broadened absorption. When the pump power is 300\,mW, the observed conical emission with a pattern of ring is shown in Fig.\,S2(a). The ring has a half-angle cone of rough 8\,mrad. The conical emission angle is affected by the system parameters, such as the cell temperature and the pump's one-photon detuning frequency\,\cite{PhaseMatch}. 

After filtering out the residual pump power with a Golan-Taylor polarizer after the cell, the ring contains photons at the probe and conjugate frequencies with a frequency difference of $\sim$6\,GHz (Fig.\,S2(c)). The majority of the generated photons are at the conjugate frequency due to the strong absorption at the probe frequency with the present rubidium system\,(Fig.\,S1). When the system is injected with 1\,pW continuous input power, the observed pattern is shown in Fig.\,S2(b). The probe beam's intensity is weaker due to the additional Doppler-broadened absorption.

\section{Group-velocity delay times versus pump power}

In the present experiment, the group-velocity delay times can be tuned by changing the pump power. Due to the power broadening, the gain profile's bandwidth is proportional to the pump intensity\,(Fig.\,S3 inset), which modifies the dispersion to be less steep as the pump power increases, resulting in less group-velocity delay. Fig.\,S3 shows the output pulse delay's dependence on the pump power when the input probe pulses contain 3.8 photons on average. The measurement in Fig.\,S3 shows that the group delay can be tuned by changing the pump power. The relative delay between the conjugate and probe pulse is a fundamental feature of this scheme\,\cite{FWMSlow, LukinScience2003, Lukin2000}.

\bigskip
\bigskip
\pagebreak
\newpage

\end{document}